\documentclass[conference]{IEEEtran}
\IEEEoverridecommandlockouts
\usepackage{cite}
\usepackage{amsmath,amssymb,amsfonts}
\usepackage{algorithmic}
\usepackage{graphicx}
\usepackage{multirow}
\usepackage{textcomp}
\usepackage{tikz}
\usetikzlibrary{positioning,arrows.meta}
\usetikzlibrary{shapes,arrows,positioning}
\usepackage{xcolor}
\def\BibTeX{{\rm B\kern-.05em{\sc i\kern-.025em b}\kern-.08em
    T\kern-.1667em\lower.7ex\hbox{E}\kern-.125emX}}

\begin{document}
\title{Exponential Weighting Model Predictive Control with Observer for Modular Multilevel Converters\\
{\footnotesize \textsuperscript{}}
}
\author{\centering
\IEEEauthorblockN{Sunny Singh}
\IEEEauthorblockA{\textit{TU Delft},
Delft, Netherlands \\ s.singh-6@tudelft.nl}
\and
\IEEEauthorblockN{Saurabh Mishra}
\IEEEauthorblockA{\textit{IIT (BHU)},
Varanasi, India \\
 saurabhmishra.rs.mat23@itbhu.ac.in}
\and 
\IEEEauthorblockN{Dušan M Stipanović}
\IEEEauthorblockA{\textit{University of Illinois},
Urbana, IL, USA \\
dusan@illinois.edu} \and 
\IEEEauthorblockN{Aleksandra Lekić}
\IEEEauthorblockA{\textit{TU Delft},
Delft, Netherlands \\
a.lekic@tudelft.nl}
}

\maketitle
\begin{abstract}
In this article, we propose a model predictive control (MPC) scheme with an exponential cost function, along with an observer for the Modular Multilevel Converter (MMC), to enhance converter dynamic performance.
In particular, as the prediction horizon $(N_P)$ increases, the numerical conditioning deteriorates rapidly, especially when a large $N_P$ is employed. This research work uses an appropriate cost function weighted to overcome the limitations of a large $N_P$.
We further analyse the effects of constraints, observing that the designed MPC strictly adheres to them and that the control variable influences the MMC plant's response. The presence of the observer improves the prediction of the output, particularly for setpoint changes in the reference signal. We also analyze the prescribed performance, which provides a priori guarantees of closed-loop stability for the proposed controller.

\end{abstract}

\section{Introduction}
\noindent The MMC has seen widespread adoption due to its excellent scalability, modularity, and suitability for high-power and high-voltage applications, particularly in high-voltage direct current (HVDC) transmission systems~\cite{lesnicar2003innovative}. Compared to conventional converter systems, MMCs offer improved reliability, making them highly suitable for modern HVDC systems and grid integration technologies such as wind energy conversion systems and HVDC transmission systems~\cite{liu2020minimal,li2020hybrid}.
With the increasing integration of power electronic converters, the dynamic response of power systems has accelerated from the order of seconds to microseconds. Consequently, there is a growing need for robust, high-speed control strategies that deliver optimal performance while accounting for system constraints. For this, MPC \cite{shetgaonkar2023model,martin2021modulated} is the most popular, suitable, and reliable control.
 
 By appropriately shaping the cost function, multiple control objectives can be addressed within a single framework, which explains the widespread adoption of MPC in the control of multilevel converter topologies, including neutral-point-clamped, active neutral-point-clamped, MMC, cascaded $H$-bridge, and matrix converters \cite{cortes2010model,bocker2014experimental,geyer2012model, vazquez2016model}. In power electronic converters, a sufficiently long prediction horizon ($N_P$) is required to accurately capture fast switching dynamics, resonant phenomena, and the non‑minimum‑phase or delayed responses that are characteristic of these systems. If the $N_P$ is too short, it may result in instability, poor steady-state performance, or increased harmonic distortion. This occurs because the controller can only react to the initial transient, failing to anticipate how currents and voltages will evolve thereafter, which is an important sensitive parameter of the MPC. From a performance standpoint, increasing $N_p$ helps reduce harmonic distortion and the switching frequency, thereby enhancing the overall efficiency of the power converter \cite{geyer2016model,zafra2023long,karamanakos2019guidelines}.
A further motivation is the numerical sensitivity issues arising from the Hessian matrix's increasing condition number when using large $N_P$ in MPC, leading to predicted outputs that vary across different horizons. 
To achieve lower control effort, reduced parameter sensitivity, and a smaller condition number of the Hessian matrix simultaneously, we adopt an exponential-based MPC technique that satisfies all these requirements.
Further motivation stems from integrating a Luenberger observer~\cite{luenberger2003observers,chakraborty2021capacitor} within the exponential-based MPC framework. At each time step, the observer introduces a correction term based on discrepancies between measured states and model predictions, thereby compensating for modeling inaccuracies and disturbances.
This approach reduces the controller's sensitivity to $N_P$ and is equivalent to a discrete linear-quadratic regulator (DLQR) formulation with an infinite $N_P$, and it also handles constraints. 
To date, no work has combined exponential-based MPC with state observers to analyze MMC performance for sufficiently large $N_P$; this research gap motivates our study, which compares the proposed controller against both DLQR and non-observer-based MPC approaches.

This paper presents the use of the exponential-based MPC for the MMC to overcome the issue with a long prediction horizon ($N_P$). Moreover, the results are compared with the DLQR and without observer-based MPC, and it is shown that the proposed controller outperforms. Furthermore, a state observer enhances the prediction accuracy, reduces setpoint tracking error of the predicted output currents, and improves the robustness of MMC plant performance under both small and large disturbances. Lastly, this paper investigates the prescribed stability, ensuring that the plant's eigenvalues lie within prescribed bounds. Performance indices of the proposed controller are also evaluated against non-observer-based and non-exponential controllers.

\section{MMC and MPC Formulation}
\noindent In this work, we consider a three-phase MMC \cite{bergna2018generalized}, described with upper and lower arm voltages and currents can be described by: $v_{Mj}^{U,L} = m_j^{U,L} v_{Cj}^{U,L}$, and $i_{Mj}^{U,L} = m_j^{U,L} i_{j}^{U,L}i_{j}^{U,L}$, for a given phase $j \in \{a, b, c\}$, and where superscripts $U$ and $L$ denote the upper and lower arms, respectively. Here, $m_j^{U, L}$  are the insertion indices for the upper and lower SMs, respectively, and $v_{Cj}^{U, L}$, $v_{Mj}^{U, L}$ represent their respective equivalent capacitor voltages. Using the $\Sigma$-$\Delta$ nomenclature, the variables in the upper and lower converter arms are given as:
\begin{align}
\nonumber i_j^\Delta &= i_j^U - i_j^L, \quad i_j^\Sigma = \frac{i_j^U + i_j^L}{2},  \quad v_{Cj}^\Delta = \frac{v_{Cj}^U - v_{Cj}^L}{2}, \\
\nonumber v_{Cj}^\Sigma &= \frac{v_{Cj}^U + v_{Cj}^L}{2}, \quad
m_j^\Delta = m_j^U - m_j^L, \quad m_j^\Sigma = m_j^U + m_j^L, \\
\nonumber v_{Mj}^\Delta &= \frac{-v_{Mj}^U+ v_{Mj}^L}{2} = \frac{-m_j^\Delta v_{Cj}^\Sigma+ m_j^\Sigma v_{Cj}^\Sigma}{2} \\
v_{Mj}^\Sigma &= \frac{v_{Mj}^U+ v_{Mj}^L}{2} = \frac{m_j^\Sigma v_{Cj}^\Sigma+ m_j^\Delta v_{Cj}^\Delta}{2}.\label{eq:arm_voltage_sigma}
\end{align}
In $dqz$-frame \cite{bergna2018generalized,shetgaonkar2021microsecond}, the differential equation describing the dynamic behavior of the MMC's output currents can be stated as:
\begin{align}
    \frac{d}{dt} \left( \vec{i}_{dq}^\Delta \right) &= \frac{\vec v_{Mdq}^\Delta-\left( -\omega L_{eq}^{ac} J_2 + R_{eq}^{ac} I_2 \right)\vec{i}_{dq}^\Delta- \vec{v}_{dq}^G}{L_{eq}^{ac}} \label{eq:differential_1},
\end{align}
where $L_{eq}^{ac} = L_{arm} + L_r/2$, $R_{eq}^{ac} = R_{arm} + R_r/2$ represent the equivalent AC resistance and inductance, respectively. The operational dynamics of the MMC  can be characterized using the AC-side modelling framework for this work \cite{zama2017high} represented by the following discrete-time differential equation: $\dot{\vec{x}}_m = A_m\vec{x}_m + B_m\vec{u}$,
where $\vec{x}_m = [ i_d^\Delta, i_q^\Delta]^T$ and $\vec{u} = [ v_{Md}^\Delta - v_d^\Delta, v_{Mq}^\Delta - v_q^\Delta]^T$, and $A_m$ and $B_m$ are matrices given by:
\begin{equation}
A_m = \begin{bmatrix}
 \frac{-R_{eq}^{ac}}{L_{eq}^{ac}} & -\omega \\
 \omega & \frac{-R_{eq}^{ac}}{L_{eq}^{ac}}
\end{bmatrix},
B_m = \begin{bmatrix}
  \frac{1}{L_{eq}^{ac}} & 0 \\
 0 & \frac{1}{L_{eq}^{ac}}
\end{bmatrix}.
\label{eq2}
\end{equation}   
We use the zero-order hold (ZOH) method to discretize the plant \cite{shetgaonkar2021microsecond,zama2017high}, as it preserves harmonic accuracy up to half the switching frequency. 
 Motivated by the \cite{zama2017high,shetgaonkar2023model}, the standard discrete state space MMC model is given with the following equation:
\begin{subequations}
\label{eq:main}
\begin{align}
\vec{x}(k + 1) = A_p(T_s)\vec{x}(k) + B_p(T_s)\vec{u}(k) \label{eq:3a} \\
\vec{y}(k+1) = C_p(T_s)\vec{x}(k), \label{eq3}
\end{align}
\end{subequations}
where $C_p(T_s)$ is an identity matrix, whereas $A_p(T_s)$ and $B_p(T_s)$ are given by $A_p(T_s) = e^{A_m T_s}$, $B_p(T_s) = \int_{0}^{T_s}e^{A_m \Tilde{T}_s}B_m d\Tilde{T}_s$. 
In this representation, $\vec{u}$ denotes the control signals while $\vec{x}_m$ represents the output signals. 
An integral term is incorporated into the output formulation to ensure zero steady-state error, yielding a discrete-time MMC model. The standard MMC given by the following equation \eqref{eq:system}
\begin{subequations}
\label{eq:system}
\begin{small}
\begin{align}
\underbrace{
\begin{bmatrix}
\Delta \vec{x}_m(k + 1) \\ 
\vec{y}(k + 1)
\end{bmatrix}
}_{\vec{x}(k+1)}
= &
\underbrace{
\begin{bmatrix}
A_p(T_s) & \mathbf{0}^T \\ 
C_p(T_s)A_p(T_s) & 1
\end{bmatrix}
}_{A}
\underbrace{
\begin{bmatrix}
\Delta\vec{x}_m(k) \\ 
\vec{y}(k)
\end{bmatrix}
}_{\vec{x}(k)} + 
\nonumber\\&
\underbrace{
\begin{bmatrix}
B_p(T_s) \\ 
C_p(T_s)B_p(T_s)
\end{bmatrix}
}_{B}
\Delta\vec{u}(k),
\label{eq:state_update}
\\
\vec{y}(k) = \underbrace{[\mathbf{0}_m \quad I]}_{C}& \vec{x}(k).
\label{eq:output}
\end{align}
\end{small}
\end{subequations}

From equation~\eqref{eq:system}, it can be reformulated  
\begin{align}
x(k+1)=&Ax(k)+B\Delta u(k),\nonumber\\
y(k)=&Cx(k)\label{eq7}
\end{align}
Furthermore, $x(k)$ is reconstructed by a state observer, enabling the controller to compute an optimal input sequence even when a subset of the states is directly measurable and operational constraints are present. 

To introduce the observer in the MPC formulation, we have to add a correction term in the state space model of the MMC, and it is of the following  dynamics:
\begin{small}    
\begin{align}
&\underbrace{\hat{x}_m(k+1) = A_m \hat{x}_m(k) + B_m u(k)}_{\text{model}}
+ \underbrace{K_{ob}(y(k) - C_m \hat{x}_m(k))}_{\text{correction term}},
\label{eq9}
\end{align}
\end{small}
\noindent where $\hat{x}_k$ and $ \hat{y}_k$ denote the estimated state vector and measured system output, respectively, corresponding to the observer. Again, $K_{ob}$ is the observer gain matrix, which plays a vital role in determining the convergence rate of the estimation error and can be adjusted as desired. 
% Then we can use this model to calculate the state variable $\hat{x}_m(k)$, $k = 1, 2, \ldots$, with an initial state condition $\hat{x}_m(0)$ and input signal $u(k)$.
And, the corresponding output becomes $\hat{y}$ as:
\begin{align}
\hat{y}=C\hat{x}(k).  \label{eq10}
\end{align}
Then, based on the state  $\hat{x}(k_i)$, the future state are calculated over the $N_p$ iteratively as: 
\begin{align}
\hat{x}(k_i+N_p|k_i) = A^{N_p} \hat{x}(k_i) + \sum_{i=0}^{N_c - 1} A^{N_p - i - 1} B \Delta u(k_i+i),
\label{eq10a}
\end{align}
where $k_i$ is the current time, and $N_c$ is the control horizon. 
The term $k_i + N_p \,|\, k_i$ denotes the prediction of the state at time $k_i + N_p$, provided time $k_i$.  The corresponding predicted output is:
\begin{small}
\begin{align}
&\hat{y}(k_i+N_p|k_i) = C A^{N_p} \hat{x}(k_i) + \sum_{i=0}^{N_c - 1} C A^{N_p - i - 1} B \Delta u(k_i+i)
\label{eq10b}
\end{align}    
\end{small}
The  $\Delta u$ over time in the MPC context can be approximated using a finite set of discrete-time Laguerre functions \cite{wang2009model}. This approach simplifies the optimization problem by reducing the number of decision variables and placing a lighter computational burden.
A set of Laguerre functions can determine the control trajectory as:
\begin{align}
\Delta u(k_{i}+k)=\sum_{j=1}^{N} (k_{i}) s_{j}(i)=S(k)^{T} \eta  \label{eq12}
\end{align}
Here, $S(k) = [s_1(k),\, s_2(k),\, \dots,\, s_N(k)]^T$ is the discrete-time Laguerre functions, and ${\eta = [c_1,\, c_2,\, \dots,\, c_N]^T}$ is the corresponding coefficient vector. The parameter $N$ denotes the number of terms used in the trajectory approximation, where $s_j(i)$ denotes its $j$-th Laguerre function and $c_j$ its associated coefficient, for $j \in \{1, 2, \dots \}$.
 The $S(k)$ function is calculated iteratively as:
\begin{align}
\nonumber &\mathbf{S}(k+1)=A_{l} \mathbf{S}(k), \label{eq13}\\
\nonumber &A_{s}=\left[\begin{array}{ccccc}a & 0  & \ldots & 0 \\ \beta & a &  \ldots & 0 \\ -a \beta & \beta &  \ldots & 0 \\ \vdots & \vdots & \vdots & \vdots\end{array}\right],
S(0)^{T}=\sqrt{\beta}\left[\begin{array}{cccccc}1,  -a &   \ldots \end{array}\right],
\end{align}    
\noindent where the factor $\beta = 1 - a^2$ depends on the pole $a$ of the discrete-time Laguerre network, and $0 \leq a < 1$. For the considered systems, each control input $\Delta u_{m}$ is linked to its own Laguerre basis $S_{m}$ and the corresponding coefficient vector $\eta_{m}$.
Then, to obtain the predicted future state using Laguerre-based methods, it is obtained by substituting \eqref{eq12} into \eqref{eq10a}:
\begin{align}
\hat{x}\left(k_{i}+N_{p} \mid k_{i}\right)=&A^{N_{p}} \hat{x}(k i)
+\sum_{i=0}^{N_{p}-1} A^{N_{p}-i-1}\bigg[B_{1} S_{1}(i)^T+\nonumber\\&B_{2} S_{2}(i)^T +\ldots+B_{m} S_{m}(i)^T\bigg] \eta_{p}, 
\end{align}
where $\eta_{p}^{T}=\left[\begin{array}{llll}\eta_{1}^{T} & \eta_{2}^{T} & \ldots & \eta_{m}^{T}\end{array}\right].$ To determine the coefficients in $\eta_p$, a constrained optimization problem must be designed and solved.
Accordingly, the control sequence is obtained by minimizing a specifically designed cost function. In this work, an exponential cost function is considered.
\section{Exponentially Weighted MPC with Observer for MMC} 
\noindent In order to determine the coefficients $\eta_p$, a constrained optimization problem must be designed and solved. In these cases, we consider the exponential-based cost-function optimisation problem with two exponential weighting factors, and the corresponding estimated control law is obtained by minimising the appropriately designed cost function. 
\subsubsection*{Case I}
In this case, the proposed exponential cost function is considered as the optimization problem in the following form:
\begin{align}
J & =\sum_{j=1}^{\infty} \lambda^{-2 j} \hat{x}^{T} Q \hat{x}   +\sum_{j=1}^{\infty} \lambda^{-2 j} \Delta u^{T} R \Delta u,\label{eq14}
\end{align}
where $\lambda > 1$ is the exponential weighting sequence.  To obtain the optimal solution by minimizing the above cost function, we can use the following transformation of variables:
\begin{eqnarray*}
&\Delta \hat{u} =\lambda^{-j} \Delta u,~~ 
X =\lambda^{-j} \hat{x}, \label{eq15a} \\
&X\left(k_{i}+j+1 \mid k_{i}\right) =\hat{A} X\left(k_{i}+j \mid k_{i}\right)+\hat{B} \Delta \hat{u}\left(k_{i}+j\right) . \label{eq15}
\end{eqnarray*}
where $\hat{A}=\lambda^{-1}.A$ and $\hat{B}=\lambda^{-1}.B$. Using these transformations, equation \eqref{eq14} is reformulated in terms of transformed variables.
The optimal solution of \eqref{eq14} is computed using this transformation, subject to constraints on the control inputs, where
$u_{\min } \leq M \eta+u(k-1) \leq u_{\max }.$
\begin{small}
\begin{align*}
 \begin{gathered}
M=\left[\begin{array}{cccc}
\sum_{i=0}^{k-1} S_{1}(i)^{T} & \mathbf{0}_{2}^{T} & \cdots & \mathbf{0}_{m}^{T} \\
\mathbf{0}_{1}^{T} & \sum_{i=0}^{k-1} S_{2}(i)^{T} & \cdots & \mathbf{0}_{m}^{T} \\
\vdots & \vdots & \vdots & \vdots \\
\end{array}\right], \\
\\
\end{gathered}
  \end{align*}      
\end{small}
Then, the cost function and constraints are rewritten in terms of the transformed variables, and the optimal control input is obtained by minimizing the following cost function:
\begin{small}
\begin{align}
 J_\lambda &= \sum_{j=1}^{\infty} X_m(k_i + j \mid k_i)^T Q_\lambda X_m(k_i + j \mid k_i) + \eta_P^T R_\lambda \eta_P, \label{eq16} \end{align}
 \end{small}
\text{subject to}
\begin{small}    
\begin{align}
X_m(k_i + j+1 \mid k_i) &= \lambda^{-1} A X_m(k_i+j|k_i) +\lambda^{-1} B \, \Delta \hat{u}(k_i+j). \nonumber\\
&u_{\min } \leq M \eta + u(k-1) \leq u_{\max } \nonumber,\\
&\text{and} \nonumber \\
\gamma  =\frac{1}{\lambda}, \quad
Q_{\lambda}  =\gamma^{2} Q&+\left(1-\gamma^{2}\right) P_{\infty}, \quad
R_{\lambda}  =\gamma^{2} R, \nonumber 
\end{align}
\end{small}
 where  $Q_{\lambda} \geq 0, R_{\lambda}>0$ are weight matrices; and $u_{\min }, u_{\max }$ are the lower and upper limits in the control input, 
 and $P_{\infty}$ is obtained by the algebraic Riccati equation.

\subsubsection*{Case II Prescribed Performance of Exponential MPC along with the observer}
The design objective is to achieve the desired closed-loop behaviour specified by the decay rate parameter $\mu$. The predictive controller is set up so that all closed-loop eigenvalues of the MMC lie within a circle of radius $\mu$ in the complex plane, defined by the following form:
\begin{align}
    \|X(k_i + j \mid k_i)\| \leq \mu^j \|{X}(k_i)\|,\label{eq17}
    \end{align}
where $j$ represents the prediction step and $\mu$ has value between $0$ and $1$. Therefore, in this case, the optimal solution is obtained by the following cost function:
\begin{small}
\begin{align}
J  =&\sum_{j=1}^{\infty} \mu^{-2 j}
\hat{x}\left(k_{i}+j \mid k_{i}\right)^{T} Q_{\lambda} \hat{x}\left(k_{i}+j \mid k_{i}\right)   +\eta_P^T R_\mu \eta_P.\label{eq18}
\end{align}
\end{small}
Similar to the previous case, it employs substitution, where
 $\lambda>1, 0<\mu<1, Q \geq 0, R>0$ are given, and
is equivalent to the optimal solution of \eqref{eq16}, with the following parameters as  $\gamma =\frac{\mu}{\lambda},~ 
Q_{\lambda}  =\gamma^{2} Q+\left(1-\gamma^{2}\right) P_{\infty},~~ 
R_{\lambda} =\gamma^{2} R.$

\section{Simulation Results and discussion }\label{Sec:4}
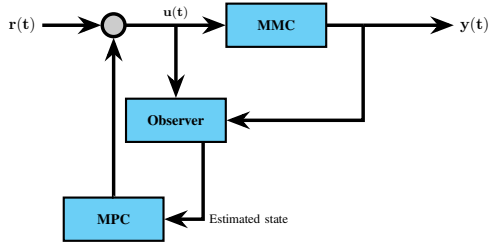
\begin{figure}[h]
    \centering
    \begin{tikzpicture}[
auto, node distance=1.2cm and 1.3cm,>=Stealth,line width=1.3pt,scale=0.6, transform shape]
% Compact styles for 2-column
\tikzstyle{block} = [draw, rectangle, minimum width=2.2cm, minimum height=0.95cm, 
 thick, fill=cyan!50, font=\small\bfseries];
% Main path nodes (more compact)
\node (r) {$\mathbf{r(t)}$};
\node[circle, draw, minimum size=5mm, right=1.3cm of r, fill=gray!40] (sum) {};
\node[block, right=2.2cm of sum] (mmc) {MMC};
    
 % Coordinates
\path (sum) -- (mmc) coordinate[pos=0.5] (midUM);
    
 % Observer and Controller (tight layout)
\node[block, below=1.6cm of midUM] (obs) {Observer};
\node[block, below=3.5cm of sum] (ctrl) {MPC};
 \node[right=2.8cm of mmc] (y) {$\mathbf{y(t)}$};

    % Connections - PERFECT 2-COLUMN LAYOUT
\draw[->] (r) -- (sum);
\draw[->] (sum) -- node[above, font=\footnotesize] {$\mathbf{u(t)}$} (mmc);
    \draw[->] (midUM) -- (obs.north);
    \draw[<-] (sum.south) -- ++(0,-0.15) |- (ctrl.north);
  % Output paths
\path (mmc.east) -- ++(0.8,0) coordinate (midY);
    \draw[-] (mmc.east) -- (midY);
    \draw[->] (midY) -- (y);
    \draw[->] (midY) |- (obs.east);
    \draw[->] ([xshift=6mm]obs.south) -- ++(0,-0.6) |- 
    node[right,pos=0.5, font=\footnotesize] {Estimated state} (ctrl.east);
\end{tikzpicture}
\caption{ MPC with observer-based feedback.}
 \label{fig:mmc_observer}
\end{figure}

To evaluate the performance of the MMC controlled by exponential and observer-based MPC, the parameters of the plant are given in Table \ref{tab:mmc_params}. In order to observe the behavior of the output current in the $dq$ frame, we have considered several (short and long) disturbances to test the performance of the controller, which are also given in Table \ref{tab:mpc_cases}. The operation of the proposed controller is depicted in the block diagram shown in Fig. \ref{fig:mmc_observer}. Again, the MPC parameters considered for the simulations are \( Q = C^{\top} C \), \( R = 1 \times 10^{-5} I_2 \), \( a = [0.4, 0.4] \),
and \( N_p \in \{50, 90, \dots, 340\} \).

 \vspace*{-5mm}
\begin{table}[h]
\centering
\caption{System Parameters of MMC in [p.u.]}
\label{tab:mmc_params}
\begin{tabular}{|l|l|l|}
\hline
\textbf{Name} & \textbf{Notation} & \textbf{Numeric Value} \\ \hline
Inductance & \( L_{arm} \) & 0.1500  \\ \hline
Resistance & \( R_{arm} \) & 0.0015  \\ \hline
AC filter inductance & \( L_f \) & 0.1200  \\ \hline
AC filter resistance & \( R_f \) & 0.003  \\ \hline
Sample time & \( T_s \) & 2.00 [ms] \\ \hline
\end{tabular}
\vspace*{-5mm}
\end{table}
\begin{small} 
\begin{table}[h]
\centering
\caption{Simulation scenarios for observer-based MPC under small and large disturbances.}
\label{tab:mpc_cases}
\renewcommand{\arraystretch}{1.1}
\begin{tabular}{|c|c|c|c|}
\hline
\textbf{Changes} & \textbf{Variable} & \textbf{1st Change} & \textbf{2nd Change} \\
\hline

\multirow{2}{*}{Small} 
& Active power & -0.500 p.u. (10) & -0.5 p.u. (55)\\
& Reactive Power & 0.500 p.u. (10) & 0.5 p.u. (55)  \\
\hline

\multirow{2}{*}{Large} 
& Active power & \multicolumn{2}{c|}{1.00 p.u. (40)} \\
& Reactive Power & \multicolumn{2}{c|}{-1.00 p.u. (40)} \\
\hline

\end{tabular}
\end{table}   
\end{small}
%block diagram of the proposed controller

\subsubsection{Simulation results and discussion for large $N_P$ along with the observer for Case I}
\begin{figure}[h]
    \centering
       \includegraphics[width=\linewidth]{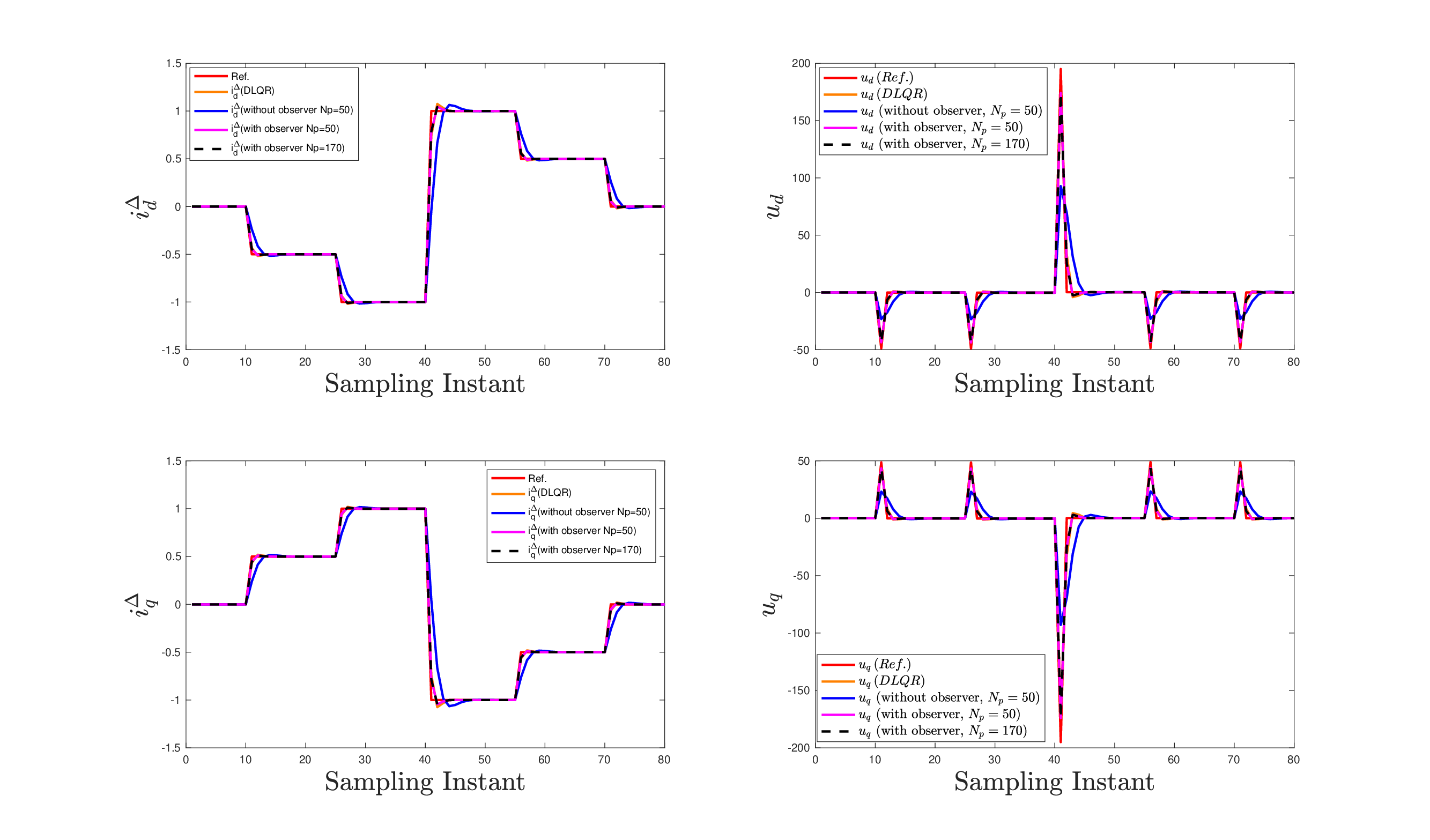}
    \caption{ $i_d^{\Delta}$, $i_q^{\Delta}$ currents and corresponding $u_d$, $u_q$ voltages for different $ N_P$.}
    \label{fig2a}
      \includegraphics[width=\linewidth]{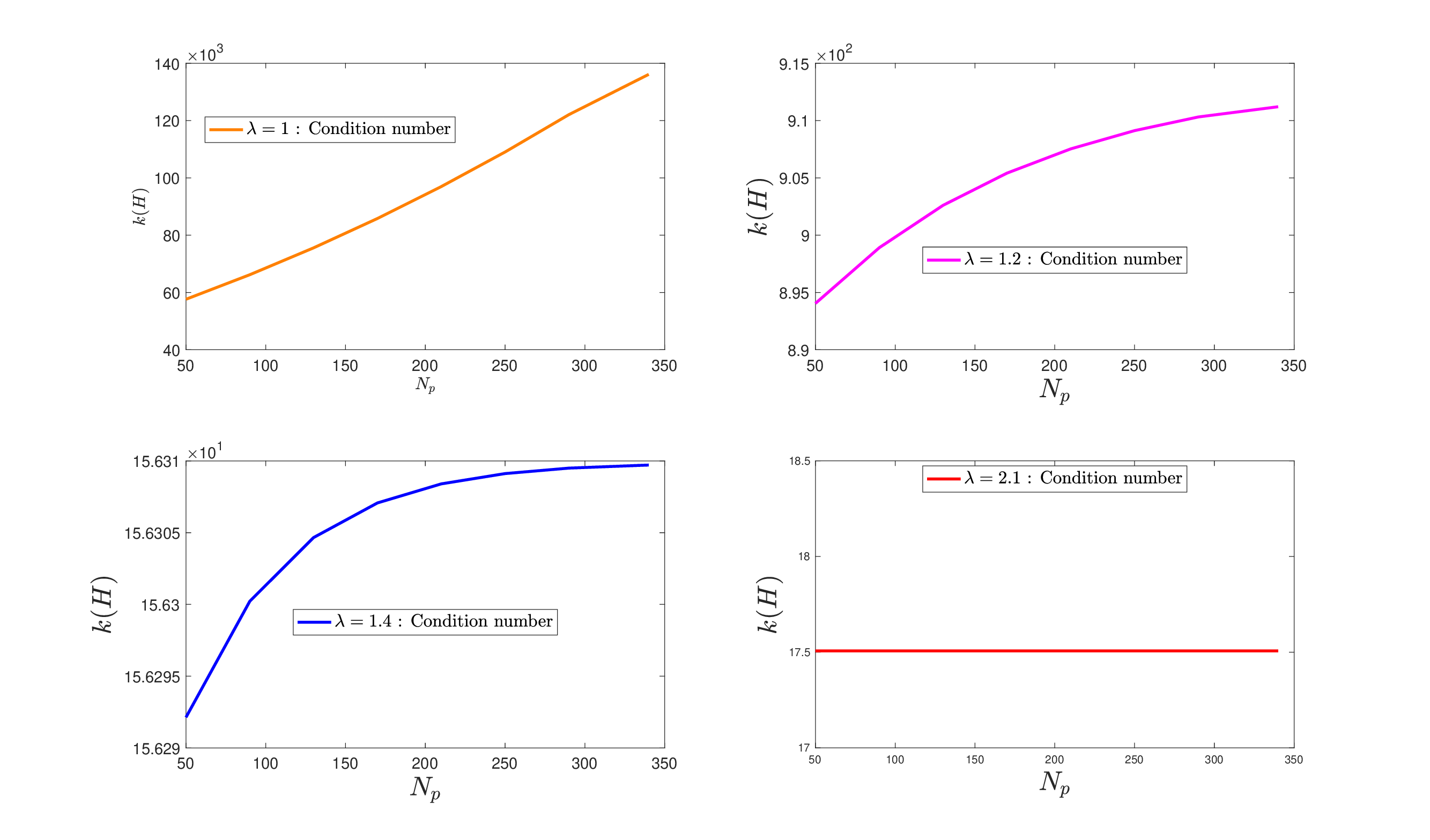}
    \caption{Condition number $\kappa(H)$ with different values of $ \lambda.$  } 
    \label{fig2c}
    \vspace*{-5mm}
\end{figure}

% Hessian Matrix
The exponential weighting with observer data reveals that, even with large $N_P$, the exponential and observer-based MPC yields a much lower condition number of the Hessian matrix than the non-exponential and non-observer case, as shown in Table \ref{tabth2}. When the factor $\lambda$ is greater than unity, the condition number rapidly decreases and converges to a finite value across a wide range of horizons, indicating improved numerical stability as shown in Fig. \ref{fig2c}. 
\begin{table}[h!]
\centering
\caption{$\kappa(H)$ with different exponential $\lambda$ and large $N_P$ Values}
\begin{tabular}{|c|c|c|c|}
\hline
\textbf{$N_p$} & \textbf{$\kappa(H)$ for $\lambda=1$} & \textbf{$\kappa(H)$ for $\lambda=1.2$} & \textbf{$\kappa(H)$ for $\lambda=1.4$} \\
\hline
50  & $5.762\times10^{4}$   & 894.0430 & 156.2921  \\
90  & $6.618\times10^{4}$  & 898.9029 & 156.3002\\
170 & $8.582\times10^{4}$ & 905.4107 & 156.3071  \\
290 & $1.220\times10^{5}$& 910.3148 & 156.3095\\

\hline
\end{tabular}
\label{tabth2}
\end{table}
In this case, we investigate the dynamic performance of the proposed MPC. To this end, we set the setpoint for both short and large disturbances to test the controller's performance, and we use DLQR as the basis for the proposed controller in the unconstrained case.
At the initial stage of the simulation, the active and reactive power are set to $0$. As the setpoint signals vary over time, the proposed MPC response closely follows the MMC closed-loop step response. For small perturbations in the setpoint, the overshoot in the $dq$-frame output currents remains negligible. When a larger disturbance is applied, the overshoot increases relative to the small-disturbance case.

The proposed MPC yields a faster response with lesser setpoint tracking errors than the MPC without an observer, as depicted in Fig.~\ref{fig2a}. The impact of the large disturbance in the active and reactive power reversal from $-1 $ to $1$ p.u. at the $40$th sampling instant is reflected in $u_d$ and $u_q$, the associated sharp peak in the MPC input, as shown in Figs.~\ref{fig2a}. Moreover, it can be observed that for larger prediction horizons $N_P$, the MPC output response remains essentially identical, while the condition number is reduced and the dynamic performance is enhanced in the presence of the observer, and it matches the DLQR response.
Again, the simulated observer gain is: 
\[K_{\text{ob}} =
\begin{bmatrix}
 0.8521 & 0.0018 & -0.0018 & 0.8521 \\
 1.8140 & 0.0018 & -0.0018 & 1.8140
\end{bmatrix}^{\top},\] and the performance indices IAE, ISE, and RMSE for this case are
\(1.7507\), \(0.32219\), and \(0.04487\), respectively, for \(N_P = 90\).
For \(N_P = 170\) and larger $N_P$, the proposed controller
yields almost identical values of these performance indices, indicating
that further increasing \(N_P\) does not lead to a significant improvement
whereas for the non-observer case, performance indices IAE, ISE, and RMSE are
\(5.9829\), \(2.8739\), and \(0.13402\), respectively, for \(N_P = 90\).
The total control effort for the proposed scheme is $0.6777$ for the prediction horizon $N_P=90$, and it is also the same for the larger prediction horizon. In closed-loop performance, the proposed controller guarantees a much larger prediction horizon, robustness, and better prediction performance.
%for the Case-II
\begin{table}[h!]
\centering
\caption{Condition number $\kappa(H)$ with different exponential factors $\lambda$, $\mu$ and large $N_p$ values}
\begin{tabular}{|c|c|c|c|}
\hline
\textbf{$N_p$} & \textbf{$\lambda = 1.1,\ \mu = 0.8$} & \textbf{$\lambda = 1.2,\ \mu = 0.8$} & \textbf{$\lambda = 2.1,\ \mu = 0.8$} \\
\hline
50  & $7.1253 \times 10^{3}$ & $1.7223\times 10^{3}$ & 71.3810 \\
90  & $5.3476 \times 10^{3}$ & $1.2636\times 10^{3}$ & 58.2711 \\
170 & $3.0070 \times 10^{3}$ & $9.274\times 10^{2}$ & 38.2943 \\
290 & $1.3288 \times 10^{3}$ &$2.906\times 10^{2}$ & 20.7934 \\
340 & $1.0311 \times 10^{3}$ &$2.234\times 10^{2}$& 17.1969 \\
\hline
\end{tabular}
\label{tabth3}
\end{table}

\begin{figure}[h!]
    \centering
       \includegraphics[width=\linewidth]{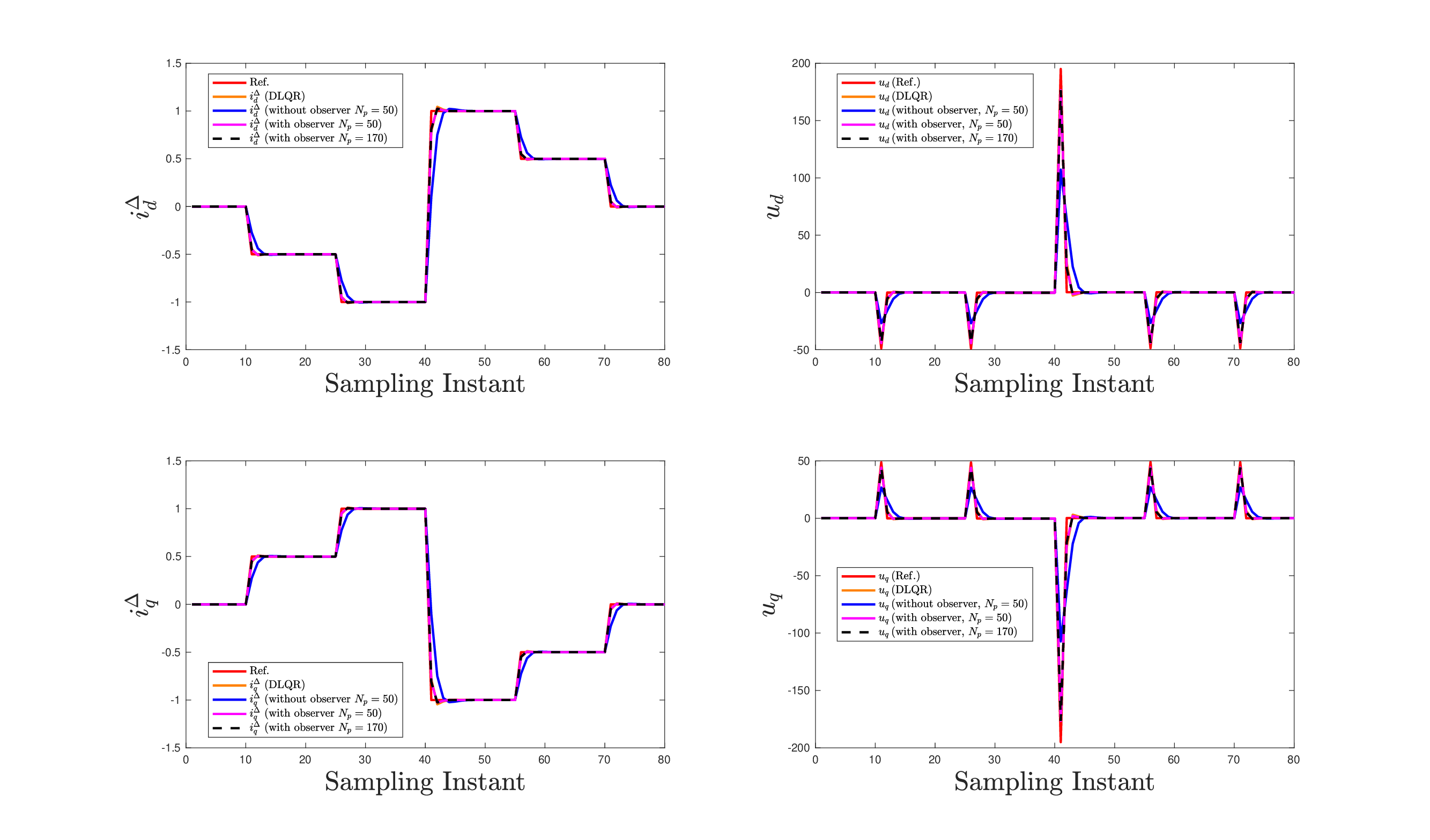}
    \caption{ $i_d^{\Delta}$, $i_q^{\Delta}$ currents and corresponding $u_d$, $u_q$ voltages for different $ N_P$.}
    \label{fig3a}
   % \end{figure}
   % \begin{figure}[h]
   %  \includegraphics[width=0.7\linewidth]{figure_all/thm3/control_input_thm3.eps}
   %  \caption{${u}_d$ voltage with and without observer and different value of $ N_p$.} \label{fig3b}

    \end{figure}
   \begin{figure}[h!]
    \includegraphics[width=\linewidth]{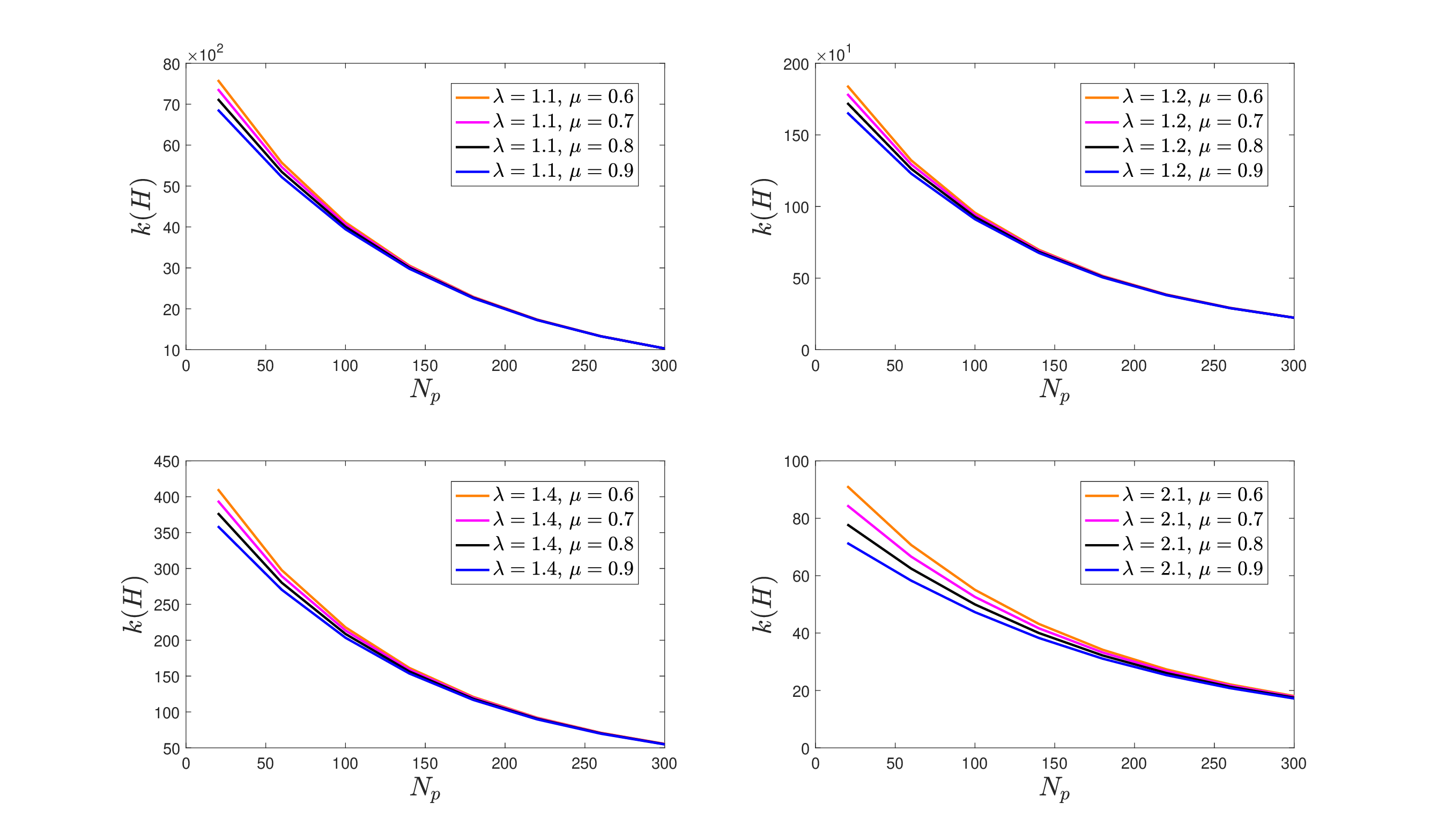}
    \caption{Condition number $\kappa(H)$ with different values of $ \lambda$. } 
    \label{fig3c}
        \includegraphics[width=\linewidth]{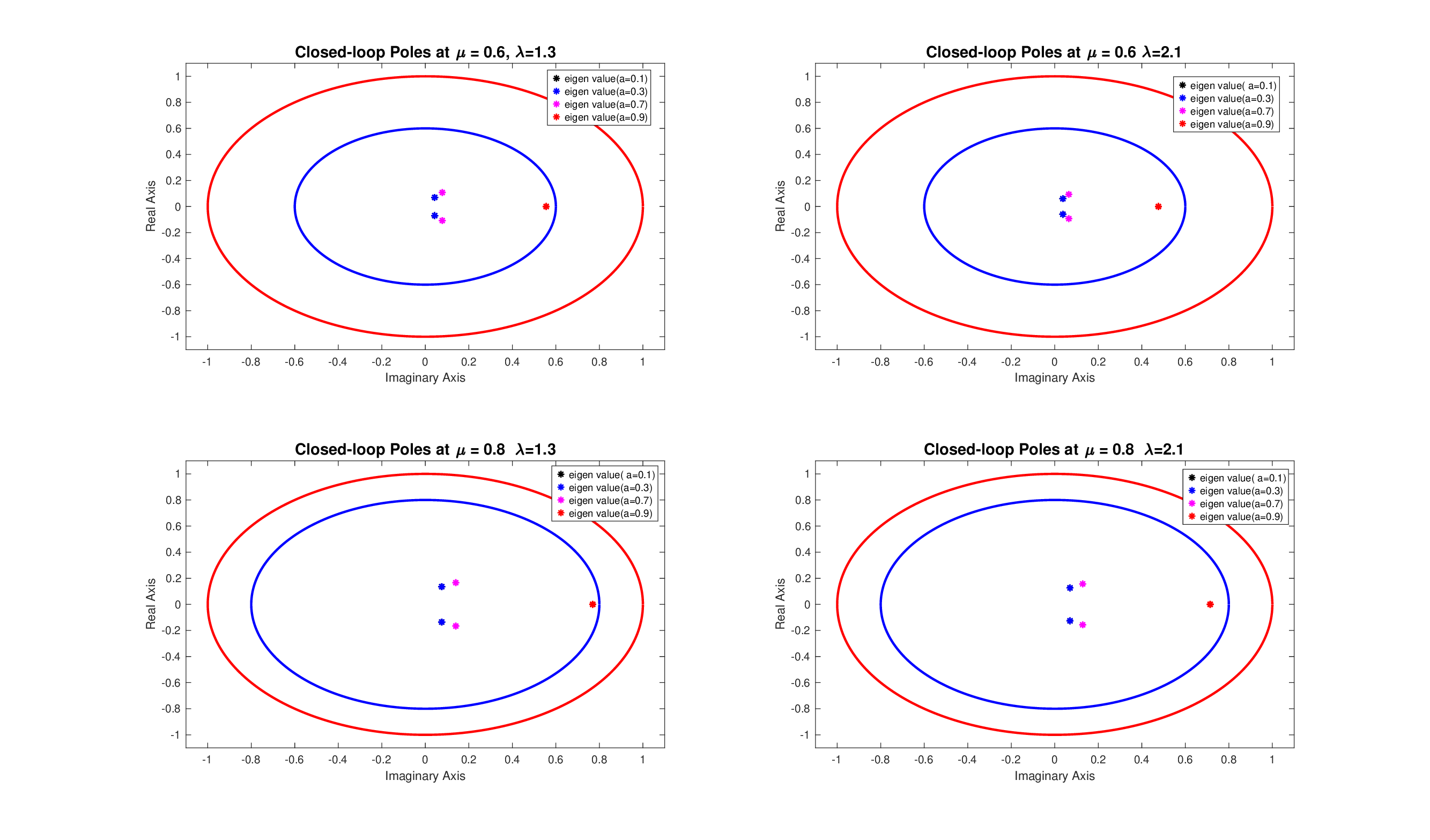}
    \caption{Closed loop eigen values with different values of the exponential factor.} 
    \label{fig3d}
       \centering
       \includegraphics[width=\linewidth]{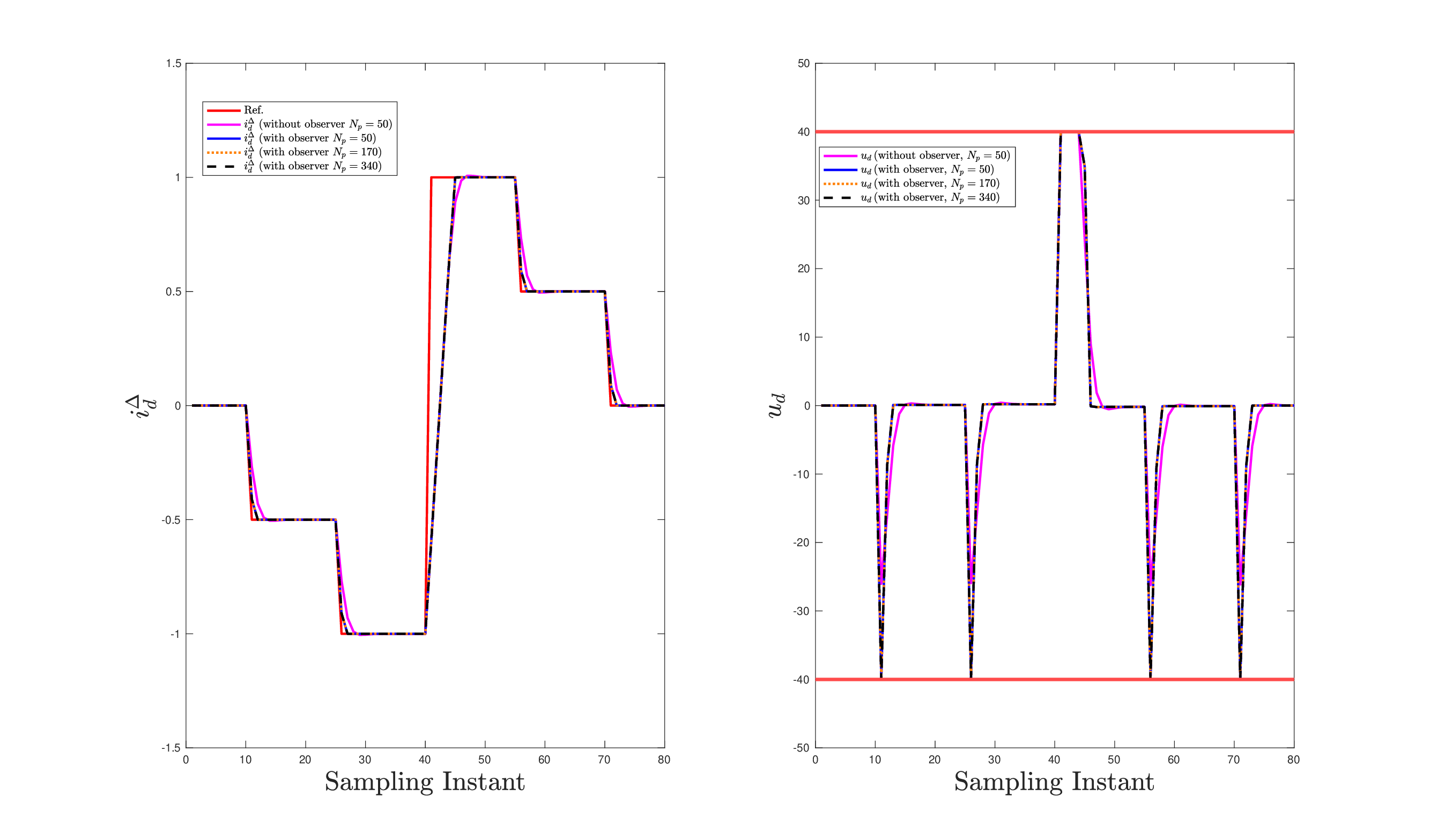}
    \caption{ Constrained based $i_d^{\Delta}$ current with and without observer with different $N_P$.}
    \label{fig3e}
    \vspace*{-5mm}
%    \end{figure}
%    \begin{figure}[h]
%     \includegraphics[width=0.8\linewidth]{figure_all/constraint/control_input_cons_thm3.eps}
%     \caption{${u}_d$ voltage with and without observer and different value of $ N_p$.} \label{fig3f}
     \end{figure}
\begin{figure}[h!]
    \centering
       \includegraphics[width=\linewidth]{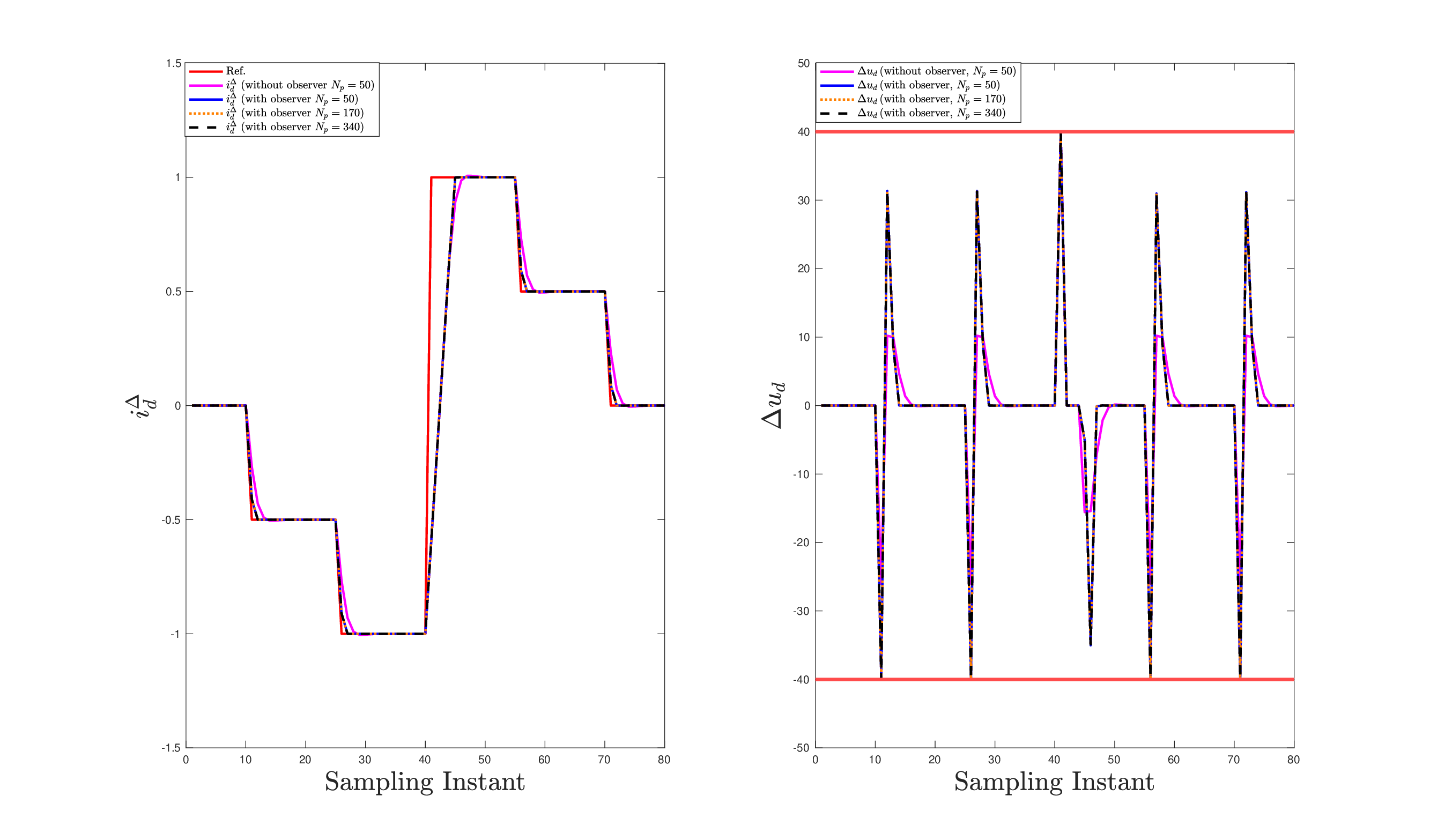}
    \caption{Constrained based $i_d^{\Delta}$ current with and without observer with different $N_P$.}
    \label{fig3g}\
    \vspace*{-5mm}
   \end{figure}

Again, the simulation for Case II, Table \ref{tabth3}, clearly shows that as $N_P$ increases, the condition number $\kappa(H)$ converges to bounded, finite values across different exponential factors $\lambda$ and $\mu$. The system exhibits remarkable numerical stability; even with minor adjustments to the exponential factors, the condition numbers remain strictly within a two to three-digit range, as shown in Fig. \ref{fig3c}. Conversely, the standard non-exponential and non-observer configurations yield poorly conditioned $\kappa(H)$, with condition numbers exceeding $10^5$. 
Again, for this case, the performance indices of the proposed controller differ significantly from those in the first case, and  IAE, ISE, and RMSE are
\(1.4038\), \(0.24364\), and \(0.039022\), respectively, for \(N_P = 90\).
For \(N_P = 170\) and larger $N_P$, the proposed controller
yields almost identical values of these performance indices, indicating
that further increasing \(N_P\) does not lead to a significant improvement in this case as well. The simulated observer gain is \[
K_{\text{ob}} =
\begin{bmatrix}
 0.8731 & 0.0019 & -0.0018 & 0.8731 \\
 1.8490 & 0.0019 & -0.0018 & 1.8490
\end{bmatrix}^{\top}.
\]

Fig.~\ref{fig3a} clearly demonstrates that the proposed observer-based MPC controller significantly outperforms the non-observer case, achieving reduced setpoint error and overshoot even under large disturbances with varying $N_p$. Notably, the closed-loop response remains robust across different $N_p$ values. Unlike the non-exponential case, which exhibits poor setpoint tracking, larger overshoot, and significantly larger $\kappa(H)$, and different responses for different $N_P$.
Moreover, for reference, DLQR is also plotted, and the response of our controller matches that of DLQR, demonstrating the significant advantages of the proposed controller. 
Again, a clear representation of the closed-loop eigenvalues under different prescribed degrees of stability is presented in Fig.~\ref{fig3d}. The results show that the closed-loop eigenvalues lie within the prescribed stability region. In particular, when a sufficiently small prescribed degree of stability is selected, the closed-loop eigenvalues remain well inside the corresponding circle. This highlights an important advantage of the proposed approach by prescribing the degree of stability in advance; stable closed-loop performance can be systematically ensured. 
Further investigate the effect of the amplitude constraint $|u_d| = |u_q| = 40$, and we compare the closed-loop response. When this amplitude limit is imposed, the closed-loop performance is slightly degraded compared to the unconstrained scenario; however, the system still exhibits satisfactory dynamic behaviour. The analysis shows that, for small disturbances, the constraints are not violated, whereas for larger disturbances, the optimal control action approaches the bounds but remains within the prescribed limits due to the optimization mechanism. Consequently, the corresponding current dynamics in the $dq$ frame become slightly more conservative than in the unconstrained case. Even under these input constraints, the proposed controller still outperforms, as clearly illustrated in Fig.~\ref{fig3e}.
 A similar behaviour is observed when the rate constraint $|\Delta u_d| = |\Delta u_q| = 40$ is employed, and the corresponding output response is shown in Fig.~\ref{fig3g}. From this, it can be clearly concluded that the proposed observer-based MPC outperforms the non-observer-based MPC in the rate-constrained scenarios. In particular, the proposed controller respects the imposed limits without violating the rate constraints, while effectively reducing the set-point tracking error of the output currents in the $dq$ frame. Output and its corresponding rate constraint are depicted in Fig. \ref{fig3g}.
 \section{Conclusion}\label{Sec:5}
\noindent We investigate the dynamic performance of the most demanding power converters using an exponential and observer-based MPC, which enhances their dynamic performance.
By using this control strategy, we addressed the long prediction horizon with fewer conditioning numbers and lower parameter sensitivities. Including the observer also reduces the error between the set point and the output prediction. In this work, constraints are also taken into account, and the corresponding output current predication of the plant in the $dq$ frame is discussed. We compare our results with non-observer-based MPC and again discuss the performance indices and control effort over the nonexponential and other controllers. Again, the prescribed closed-loop eigenvalues ensure both stability and the desired decay rate in advance. 
%Future work will explore data-driven MPC for the nonlinear MMC model. 
\bibliographystyle{ieeetr}
\bibliography{cas-ref}

\end{document}